\documentclass[USenglish]{article}	

\usepackage[utf8]{inputenc}				
\usepackage[big,online]{dgruyter}	
\usepackage{lmodern} 
\usepackage{microtype}
\usepackage[numbers,square,sort&compress]{natbib}
\usepackage{xcolor}

\theoremstyle{dgthm}

\newtheorem{lemma}{Lemma}

\theoremstyle{dgdef}
\newtheorem{definition}{Definition}

\usepackage{color,xcolor,soul}

\begin{document}

	\articletype{Research Article}

\title{A New  Hybrid Cryptosystem Involving DNA, Rabin,  One Time Pad and Fiestel}


\author*[1]{Benatmane Sara}
\author[2]{Nuh Aydin}
\author[3]{Behloul Djilali}
\author[4]{Prokash Barman}

\runningauthor{S Benatmane et al.}
\affil[1]{\protect\raggedright 
USTHB, Department of Mathematics, Bab ezzouar, Algeria, e-mail: benatmanesara34@gmail.com}
\affil[2]{\protect\raggedright 
Department of Mathematics and Statistics, Kenyon College e-mail: aydinn@kenyon.edu }
\affil[3]{\protect\raggedright 
USTHB, Department of Mathematics, Bab ezzouar, Algeria, e-mail: dbehloul@yahoo.fr}
\affil[4]{\protect\raggedright 
University of Calcutta, Department of Computer Science and Engineering, West Bengal, India, e-mail: pbcse\_rs@caluniv.ac.in}

	
\abstract{Information security is a crucial need in the modern world.  Data security is a real concern, and many customers and organizations need to protect their sensitive information from unauthorized parties and attackers. In previous years, numerous cryptographic schemes have been proposed. DNA cryptography is a  new and developing field that combines the computational and biological worlds. DNA cryptography is  intriguing due to its high storage capacity, secure data  transport, and massive parallel computing.
In this paper, a new combination is proposed that offers good security by combining DNA, the Rabin algorithm, one time pad, and a structure inspired by Fiestel. This algorithm employs two keys. The first key is a DNA OTP key which is used for only one secure communication session. The second key, which combines the public and private keys, is a Rabin key. Additionally, by using a Feistel inspired scheme and randomness provided by DNA, the ciphertext is made harder to obtain without the private key. }

\maketitle

\section{Introduction} 
 Public key cryptography is based on the notion of ``one-way function''. The security of many modern public key  cryptosystems  depends on the computational complexity of  number theoretical problems such as factorization of large integers and the discrete logarithm problem.  One of the recent methods in cryptography that can potentially increase information security is  DNA cryptography. Symmetric key cryptography uses a single key to both encrypt and decrypt the plaintext. One-time Pad, DES, and AES are a few well known symmetric key encryption schmes.
Asymmetric key encryption, also called public key cryptography (PKC), refers to cryptographic techniques that use two keys, one for encryption (known as the public key) and the other for decryption (known as the private key) (\cite{2}). Hybrid schemes are built by combining symmetric and asymmetric algorithms. Invented in $1979$ by Michael Rabin (\cite{3})  the Rabin cryptosystem  is an asymmetric cryptography algorithm similar to the RSA. The security of both cryptosystems is related to the integer factorization problem with the stronger connection in the case of Rabin system. The Rabin technique requires  simple computations in the encryption phase, whereas the decryption step can apply the Chinese Remainder Theorem to obtain the correct plaintext. Rabin proposed an RSA variation with a public key exponent of $2$ (\cite{5}). Rabin's security is more closely tied to factoring than RSA. However, it has a disadvantage for decryption.  For $N = pq$, where $p$ and $q$ are distinct primes, squaring is a four-to-one map, necessitating the use of a rule to select the proper decryption result (\cite{4}). We provide an example in our proposed algorithm for determining the correct plaintext from the four possible results.

One-time pad (OTP) is an encryption technique that was invented by Gilbert Vernam in $1917$. It is based on a much older encryption technique called "Vigenère Cipher" (\cite{1}). The OTP was improved upon by American general Joseph Mauborgne who used the random characters of the encryption key which encrypts a message with a key by adding the key to each part of the message one by one. The key can only be used once.

DNA cryptography uses DNA sequences to hide data which can be done utilizing various DNA technologies and biological techniques. Data hiding techniques have grown in popularity as a means of sending covert information in recent years. To prevent malicious intrusion and ensure a secure transmission, methods for data hiding based on DNA sequences have received a lot of attention. To address the issue of data hiding, numerous cryptographic techniques based on DNA were presented such as the method suggested in (\cite{6}) which combines genetic engineering with cryptology technology to create an asymmetric encryption and signature cryptosystem. It is an investigation into biological cryptology. DNA-PKC uses two pairs of keys for encryption and signing, similar to the traditional public key cryptography. The technique suggested in (\cite{7}) presents a strategy based on DNA and the RSA cryptosystem capable of providing an architectural foundation for the encryption and generation of digital signatures for all characters, simple text data, and text files. The process is broken down into four steps: key generation, data pre/post processing, DNA, and signature generation. The technique suggested in (\cite{8}) proposes a new cryptography technique using symmetric key exchange, OTP and DNA hybridization operation. The XOR operation with OTP DNA sequence is used as an encryption technique. The proposed technique used the matrix of the message and OTP DNA sequence to minimize the  time complexity of encryption and decryption. The conclusion is that DNA cryptography could be combined with traditional cryptography to create a hybrid system with a higher level of security. The technique  given in (\cite{9}) provides an enhanced cryptographic strategy based on DNA cryptography employing a Feistel inspired structure which is based on symmetric key cryptography. The method presented in (\cite{10}) creates an improved cryptographic algorithm. By combining current symmetric and asymmetric encryption, the proposed methodology makes use of the strengths of various popular symmetric and asymmetric encryption schemes: AES's S-Box mapping, which makes the encryption difficult to predict, ElGamal's key security based on the difficulty of solving discrete logarithms, chaos-based security's avalanche effect of using multiple keys, and RSA security's based on the difficulty of factorizing large numbers were all modified. The method outlined in (\cite{2}) is a new secure way for concealing data based on DNA sequences. It has two rounds of encryption which are similar to the DES algorithm an established encryption technique. The message is encrypted using two secret keys. The elliptic curve cryptography (ECC) and Gaussian kernel function (GKF) are used to create the first key, and the second key is created using an arbitrary injective mapping on the second characters repeated in the first key.

 Conceal  information in the DNA sequence has drawn a lot of interest, and much research has been done to develop a number of innovative techniques. The goals of the new DNA cryptography algorithms that have been presented are to achieve the highest level of security during data transfer and to reduce the computational time of encryption/decryption.

The primary goal of this  paper is to create a novel hybrid cryptosystem that combines the strengths of each cipher to assure good security. It attempts this by combining DNA, Rabin Algorithm, OTP and a Feistel inspired structure. The proposed system is a significant improvement over (\cite{25}) in the following ways: 
We employed the Rabin cryptosystem instead of RSA, because Rabin is more secure than RSA (\cite{5,15}), Additionally,  we added the Fiestel-inspired structure to our algorithm, which further strengthens its security.  The authors in (\cite{25}) do not discuss the details of the random key generation (generating truly random numbers/sequences is a major problem in cryptography) and state (in the abstract) that the system encrypts the binary key via an asymmetric cipher using the recipient's public key but only the asymmetric cipher is used to encrypt the plaintext. Moreover, they do not explain how the private random binary key is communicated from the sender to the receiver. They simply say ``The sender sends it to receiever''. The need to securely send this private key is one of the major drawbacks of symmetric key cryptosystems that make them impractical in many situations. Finally, the conversion of the binary encrypted text by the authors into a DNA sequence and then into a complement  of DNA provides no security benefit and only adds unnecessary computations to the encryption and decryption processes.

\section{Background Information} 
\subsection{DNA}
DNA is the genetic makeup of nearly all living things, ranging in complexity from tiny viruses to intricate humans. With two strands running counter-parallel, DNA has a double helix structure. Nucleotides are tiny, continuous polymers that make up DNA. A nitrogenous base, a sugar with five carbons, and a phosphate group are the three parts of each nucleotide. There are four distinct nucleotides, depending on the sort of nitrogenous base they contain. Adenine, Cytosine, Thiamine, and Guanine are the four distinct bases that are denoted by $A$, $C$, $T$, and $G$. All of an organism's vast and intricate information is stored in DNA using simply the letters $A$, $C$, $T$, and $G$. By creating hydrogen bonds with one another to keep the two strands of DNA connected, these bases create the structure of DNA strands. While $C$ and $G$ create bonds with one another, A makes a hydrogen bond with $T$ (\cite{11}). Before 1994, it was thought that DNA exclusively contained biological information. However, Adleman's solution to the NP-complete Hamiltonian path problem of seven vertices disproved this notion (\cite{12}). Since then, DNA has been also employed as a tool for computation (\cite{13}).
The four letters $A$, $C$, $T$, and $G$ make up the DNA language, which is used in DNA computing.
DNA's capacity for computation is now applied to cryptography as well. DNA cryptography is a promising area that, if used correctly, might put other areas of cryptography in a much tougher competitive position (\cite{14}). One of the key benefits of DNA in cryptography is its potential to generate large volumes of truly random numbers quickly for practical needs (\cite{21,24}).
\subsection{Rabin Cryptosystem}	
In $1979$, Rabin offered an alternative to RSA with public-key exponent $2$ (\cite{15}). Utilizing this exponent $2$ has the following benefits over using greater exponents:
\begin{itemize}
\item[(i)] A lighter computational burden.
\item[(ii)]  It is known that  factoring the RSA modulus $N$ is equivalent to computing square roots $\mod N$, i.e., solving (\ref{eq1}) below (\cite{5}). On the other hand, for the RSA cryptosystem while the factorization of $N$ is sufficient to break it, it is not known whether it is also necessary.  Hence, the security of the Rabin cryptosystem is more closely related to the integer factorization problem than RSA.  
\end{itemize}

However, the Rabin cryptosystem  also has some disadvantages. The main one is the fact that square roots $\mod N$ are not unique. When decyrpting a ciphertext, we need to find a way to pick the correct plaintext among the four possible candidates. Moreover, it is vulnerable to  the chosen-plain-text attack.

\begin{definition} (\cite{15}) Let $S=\mathbb{N}$ or $S=\{p \hspace{0.1cm} q: p, q \text{ are primes } \equiv 3 \hspace{0.1cm} \mod 4 \}$. The computational problem SQRT-MOD-N is: Given $N \in S$ and $y \in \mathbb{Z} / N \mathbb{Z}$ to output if $y$ is not a square modulo $N$, or a solution $x$ to $x^{2} \equiv y \hspace{0.1cm}(\bmod \hspace{0.1cm} N)$.
\end{definition}
\begin{lemma} (\cite{15}) 
SQRT-MOD- $N$ is equivalent to FACTOR.
\end{lemma}

\subsubsection{Key Generation of the Rabin Algorithm}
The Rabin algorithm key creation procedure is as follows:

\begin{itemize}
    \item[1.] Generate two very large prime numbers,  $p$ and $q$, that satisfy the conditions below:
$$
p \neq q  \hspace{1.5cm}  p \equiv q \equiv 3 \hspace{0.1cm}(\bmod 4).
$$
For example :
$$
\begin{array}{ll}
p=167 & q=127 \\
167 \text { prime number } &  127 \text { prime number } \\
167 \bmod 4=3 &  127 \text { mod } 4=3
\end{array}
$$
\item[2.] Calculate the value of $n$.
$$
\begin{aligned}
n &=p  q \\
&=167\times127 \\
&=21209.
\end{aligned}
$$
\item[3.] Publish $n$ as public key and save $p$ and $q$ as private keys.
\end{itemize}

\subsubsection{Encryption of the Rabin Algorithm}
The encryption steps of the Rabin algorithm are:
\begin{itemize}
\item[1.] Obtain the public key.
$$
n=21209.
$$
\item[2.] Let $m$ be the clear text, convert it to ASCII. Then, transform it to a binary sequence,  double the binary sequence by duplicating it, and return the binary value to decimal.
$$
\begin{aligned}
&m= H =72 \\
&m=72_{10}=1001000_{2} \\
&m=1001000 \mid 1001000_{2} \rightarrow \text { double extend } \\
&m=9288_{10}.
\end{aligned}
$$
\item[3.] Encrypt $m$  by squaring: \begin{equation} \label{eq1}
C =m^{2} \mod n.
\end{equation}
For the running example, we get
$$
\begin{aligned}
C &=m^{2} \mod n \\
&=9288^{2} \mod 21209. \\
&=9941.
\end{aligned}
$$
\item[4.] Send $C$ to recipient.
\end{itemize}
\subsubsection{Decryption of the Rabin Algorithm}
 With Rabin's decryption method, there are four potential outcomes, but only one of them is the correct plaintext.\\
For the running example, the steps of Rabin algorithm decryption is:
\begin{itemize}
\item [1.] Take $C$ from the sender:
$C=9941.$
\item[2.] Use the Extended Euclidean GCD algorithm to find  $y_{p}$ and $y_{q}$:
\begin{table}[h!]
    \centering
    \caption{Searching $y_{p}$ and $y_{q}$ values with the Extended Euclidean algorithm}
  \begin{tabular}{cccc}
\hline $\boldsymbol{y}_{\boldsymbol{q}}$ & $\boldsymbol{y}_{\boldsymbol{p}}$ & $\mathbf{D}$ & $\mathbf{k}$ \\
\hline 0 & 1 & 167 & \\
\hline 1 & 0 & 127 & 1 \\
\hline$-1$ & 1 & 40 &3 \\
\hline 4 & $-3$ & 7 & 5 \\
\hline -21 & $16$ & 5 & 1\\
\hline 25 & $-19$ & 2 & 2\\
\hline -71 & $54$ & 1 & \\
\hline
\end{tabular}
    \label{tab:my_label}
\end{table}\\
$y_{p}=54$  and  $y_{q}=-71.$
\item[3.] Calculate $m_{p}$ and $m_{q}$
$$
\begin{array}{lllll}
m_{p}=C^{\frac{p+1}{4}} \bmod p & & & & m_{q}=C^{\frac{q+1}{4}} \bmod q \\
m_{p}=9941^{\frac{167+1}{4}} \bmod 167 & & & & m_{q}=9941^{\frac{127+1}{4}} \bmod 127 \\
m_{p}=64 & & & & m_{q}=17
\end{array}
$$

\item[4.] Calculate the values of $r, s, t$ and $u$ using Chinese Remainder Theorem (CRT).
\begin{itemize}

\item[$\bullet$] Calculate the  values of the variables $v$ and $w$ first to facilitate the calculation.\\
$v=y_{p} \times p \times m_{q} =54\times 167 \times
17 \hspace{0.1cm}= \hspace{0.1cm}153306.$\\
$w=y_{q} \times q \times m_{p}= -71 \times 127\times64= -577088$.

\item[$\bullet$] Calculate the values of $\mathrm{r}, \mathrm{s}, \mathrm{t}$ and $\mathrm{u}$.
$$
\begin{array}{ll}
r=(v+w) \bmod n & =398, \\
s=(v-w) \bmod n & =11921, \\
t=(-v+w) \bmod n & =9288, \\
u=(-v-w) \bmod n & =20811.
\end{array}
$$
\end{itemize}
\item[5.] Convert $r, s, t$ and $u$ into binary.\\
$r=398_{10}=\quad 110001110_{2},$\\
$s=11921_{10}=10111010010001_{2},$\\
$t=9288_{10}= 10010001001000_{2},$\\
$u=20811_{10}=101000101001011_{2}.$
\item[6.] Determine the original message from the values of $r, s, t$ and $u$.
 \begin{table}[h!]
     \centering
     \caption{Determination of Original Messages}
     \begin{tabular}{crrc}
\hline$r$ & $398_{10}$ & $110001110_{2}$ & digits number is odd \\
\hline$s$ & $11921_{10}$ & $10111010010001_{2}$ & digits number is odd \\
\hline$t$ & $9288_{10}$ & $1001000\mid1001000_{2}$ & left $=$ right \\
\hline$u$ & $20811_{10}$ & $ 101000101001011_{2}$ & digits number is odd \\
\hline 
\end{tabular}

     \label{tab:my_label}
 \end{table}
 \\
 The message is in $t.$ \\
 $m=1001000_{2}=72_{10}=H.$
 \end{itemize}

\section{The Proposed  Algorithm}
The proposed algorithm is a combination of Rabin’s cryptosystem, binary one-time pad, DNA cryptography, and a Feistel inspired scheme for encryption of plaintexts. This algorithm uses both symmetric and asymmetric keys.
\subsection{Key Generation}
In this proposed method, the sender generates a random DNA key, the size of the DNA  key depends on the size of the plaintext message which will be transmitted $(size \hspace{0.1cm} of \hspace{0.1cm} DNA \hspace{0.1cm} key = 26 \times size \hspace{0.1cm} of \hspace{0.1cm} plaintext)/2$. At the same time, the receiver generates a Rabin key and sends only the public key to the sender.
\subsection{Algorithm for Encryption}
In the encryption process the sender has the plaintext, the DNA key, and the Rabin public key.
The following are steps of the sender’s encryption algorithm :
\begin{itemize}
\item[\textbf{ 1)}]
 To establish the precise output when using the Rabin cryptosystem to decrypt, we perform a sort of disambiguation diagram in this stage. The sender and receiver choose a secret spy before they start the encryption process, and they modify the plaintext by placing the spy before each character of the plaintext.
\item[\textbf{2)}] Convert the random DNA key to binary via table \ref{tab:1} (called random binary key) and send it to the receiver using the method described in (\cite{16}).
\begin{table}[!h]
\centering 
\caption{Binary codes.}
\label{tab:1}
\begin{tabular}{|l|l|l|l|l|}
\hline\noalign{\smallskip}
Nucleotide&A&C&G&T\\
\hline
Code&00&01&10&11\\
\hline
\end{tabular}
\end{table}

\item[\textbf{ 3)}] Convert the new form of the message via the ASCII table.
 \item[\textbf{ 4)}] Concatenate the numerical values two by two.
  \item[\textbf{ 5)}] Apply the Rabin encryption algorithm to each ASCII value and compute the Rabin cipher.
  \item[\textbf{ 6)}] Convert each digit of the Rabin cipher to binary plaintext of $26$ digits.
 \item[\textbf{ 7)}] Perform XOR operation between binary cipher and random binary key, called XOR cipher.
 \item[\textbf{ 8)}] Reordering every $26$ digits by Feistel inspired structure, where it is divided into two equal parts ($l_0$,$r_0$), and the XOR operation is performed between the two parts, with the left part $l_1$ containing the value of $r_0$ as is, and the XOR result moving towards the second part $r_1$. The pieces $r_1$ and $l_1$ have now been concatenated.
\begin{figure}[h!]
\centering
\includegraphics[width=6cm,height=3cm]{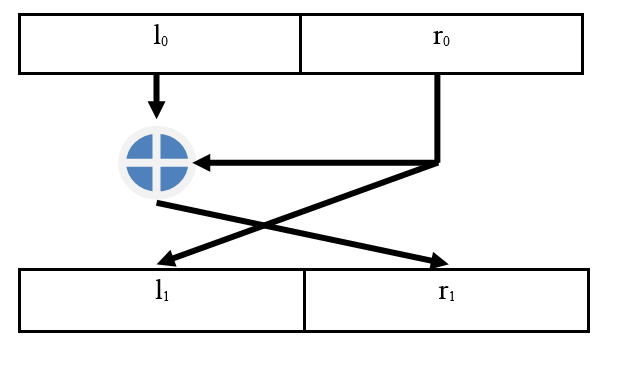}
\caption{Feistel inspired scheme for reordering binary plaintext.}
\label{fig:1}
\end{figure}
\end{itemize}
The sender sends this ciphertext to the receiver.
\subsection{Algorithm for Decryption}
The receiver's  algorithm is just opposite of sender's  algorithm.
   \begin{itemize}
 
\item[\textbf{ 1)}] Apply a concept of diffusion by using two Feistel inspired schemes for each $26$ binary digits.
\item[\textbf{ 2)}] Perform XOR operation between binary key and Feistel decryption.
\item[\textbf{ 3)}] Convert every $26$ binary cipher to corresponding decimal.
\item[\textbf{ 4)}] Decrypt each decimal by Rabin decryption key.
\item[\textbf{ 5)}]	Select the square root that starts with the ASCII character corresponding to the spy that the sender and receiver chose at the beginning from the four distinct square roots that were obtained.
\item[\textbf{ 6)}]	Remove the ASCII number associated with each selected square root's spy.
\item[\textbf{ 7)}] Convert each ASCII to its corresponding character.
   \end{itemize}
   
\section{A Didactic Example}
A prototype of this algorithm is developed by Matlab $R2016a$.\\
Sender wants to send secretly the message ``Success'' to receiver, so the size of the plaintext is $7$.
\subsection{ Generation of Key Pairs}
Alice generates a random DNA key. A DNA key is randomly selected from a website (There are millions of websites for DNA sequencing).
Randomness of DNA sequences has been utilized in several recent cryptosystems such as (\cite{22,23}). 

Size of DNA key $=$ $26 \times$ (size of plaintext)$/2$ , so the number of nucleotides is equal to $(26\times7)/2$,  DNA key is:

G T T T G G G G T T C C A A T C C A T T T A G A T C A C C C G C C G G G G T T G C C T T A C G A C A G A A T T T A T A A A C T G C A G C T T T A C A T T A C T A A T C C G C T T C T G.\\
Now DNA key is converted into binary key via table \ref{tab:1}.
\begin{table}[h!]
    \centering
    \begin{tabular}{p{0.2cm}p{0.2cm}p{0.2cm}p{0.2cm}p{0.2cm}p{0.2cm}p{0.2cm}p{0.2cm}p{0.2cm}p{0.2cm}p{0.2cm}p{0.2cm}p{0.2cm}p{0.2cm}p{0.2cm}p{0.2cm}p{0.2cm}p{0.2cm}p{0.2cm}p{0.2cm}p{0.2cm}p{0.2cm}p{0.2cm}p{0.2cm}p{0.2cm}p{0.2cm}}
1& 0& 1& 1& 1& 1& 1& 1& 1& 0& 1& 0& 1& 0& 1& 0& 1& 1& 1& 1& 0& 1& 0& 1& 0& 0\\
0& 0& 1& 1& 0& 1& 0& 1&  0& 0& 1& 1& 1& 1& 1& 1& 0& 0& 1& 0& 0&  0& 1& 1& 0& 1\\
0& 0& 0& 1& 0& 1& 0& 1& 1& 0& 0& 1& 0& 1& 1& 0& 1& 0& 1& 0& 1& 0& 1& 1& 1& 1 \\
1& 0& 0& 1& 0& 1& 1& 1& 1& 1& 0& 0& 0& 1& 1& 0& 0& 0& 0& 1& 0& 0& 1& 0& 0& 0\\
0& 0& 1& 1& 1& 1& 1& 1& 0& 0& 1& 1& 0& 0& 0& 0& 0& 0& 0& 1& 1& 1& 1& 0& 0& 1\\
0& 0& 1& 0& 0& 1& 1& 1& 1& 1& 1& 1& 0& 0& 0& 1& 0& 0& 1& 1& 1& 1& 0& 0& 0& 1\\
1 &1& 0& 0& 0& 0& 1& 1& 0& 1& 0& 1& 1& 0& 0& 1& 1& 1& 1& 1& 0& 1& 1& 1& 1& 0\\
 \end{tabular}
\end{table}
\\
The sender sends this key to the receiver using the method described in (\cite{16}). The receiver generates the Rabin key. For the running example,  the private key is $p=6911$ and $q=6947$; and  the public key is $n= 48010717$.\\
Receiver sends just $n=48010717$ as the public key to Sender. 
\subsection{Encryption}
 \begin{itemize}
\item[1)] Sender’s chooses a spy for example "$\ast$" and adds "$\ast$" at the beginning of each letter as follows:   
"$\ast$S$\ast$u$\ast$c$\ast$c$\ast$e$\ast$s$\ast$s"

\item[2)] Convert every plaintext character to corresponding ASCII character. So for this example it is:
  $$42  \hspace{0.2cm} 83 \hspace{0.2cm}   42 \hspace{0.2cm}  117  \hspace{0.2cm} 42  \hspace{0.2cm}  99 \hspace{0.2cm}   42  \hspace{0.2cm}  99 \hspace{0.2cm} 42 \hspace{0.2cm}  101  \hspace{0.2cm}  42  \hspace{0.2cm} 115  \hspace{0.2cm}  42 \hspace{0.2cm}  115.$$
  
 \item[3)]Concatenate these numerical values two by two.
$$4283    \hspace{0.2cm}  42117 \hspace{0.2cm}       4299 \hspace{0.2cm}        4299 \hspace{0.2cm}    42101 \hspace{0.2cm}       42115 \hspace{0.2cm}       42115.$$

 \item[4)] Apply Rabin encryption on each number after the concatenation. So now Rabin cipher is:\\
 \vspace{0.1cm}
 $18344089$    $45455877$    $18481401$    $18481401$ $44108389$    $45287413$    $45287413$.
 \vspace{0.1cm}
 
 \item[5)] Now every Rabin cipher is converted to $26$ digits binary value. It can be called binary cipher. 
 \begin{table}[h!]
     \centering
     \begin{tabular}{p{0.2cm}p{0.2cm}p{0.2cm}p{0.2cm}p{0.2cm}p{0.2cm}p{0.2cm}p{0.2cm}p{0.2cm}p{0.2cm}p{0.2cm}p{0.2cm}p{0.2cm}p{0.2cm}p{0.2cm}p{0.2cm}p{0.2cm}p{0.2cm}p{0.2cm}p{0.2cm}p{0.2cm}p{0.2cm}p{0.2cm}p{0.2cm}p{0.2cm}p{0.2cm}}
 0&1&0&0&0&1&0&1&1&1&1&1&1&0&1&0&0&0&1&0&0&1&1&0&0&1\\
 1&0&1&0&1&1&0&1&0&1&1&0&0&1&1&0&1&0&0&0&0&0&0&1&0&1 \\
 0&1&0&0&0&1&1&0&1&0&0&1&0&0&0&0&0&0&1&1&1&1&1&0&0&1 \\
 0&1&0&0&0&1&1&0&1&0&0&0&0&0&0&0&0&0&1&1&1&1&1&0&0&1\\
 1&0&1&0&1&0&0&0&0&1&0&0&0&0&1&0&1&0&0&1&1&0&0&1&0&1\\
 1&0&1&0&1&1&0&0&1&1&0&0&0&0&0&1&1&1&1&1&1&1&0&1&0&1\\     
 1&0&1&0&1&1&0&0&1&1&0&0&0&0&0&1&1&1&1&1&1&1&0&1&0&1\\
\end{tabular}
\end{table}
\newpage
\item[6)] Now XOR operation is performed between binary cipher and binary key: 
\begin{table}[h!]
    \centering
    \begin{tabular}{p{0.2cm}p{0.2cm}p{0.2cm}p{0.2cm}p{0.2cm}p{0.2cm}p{0.2cm}p{0.2cm}p{0.2cm}p{0.2cm}p{0.2cm}p{0.2cm}p{0.2cm}p{0.2cm}p{0.2cm}p{0.2cm}p{0.2cm}p{0.2cm}p{0.2cm}p{0.2cm}p{0.2cm}p{0.2cm}p{0.2cm}p{0.2cm}p{0.2cm}p{0.2cm}}
1&1&1&1&1&0&1&0&0&1&0&1&0&0&0&0&1&1&0&1&0&0&1&1&0&1\\
1&0&0&1&1&0&0&0&0&1&0&1&1&0&0&1&1&0&1&0&0&0&1&0&0&0\\
0&1&0&1&0&0&1&1&0&0&0&1&0&1&1&0&1&0&0&1&0&1&0&1&1&0\\
1&1&0&1&0&0&0&1&0&1&0&0&0&1&1&0&0&0&1&0&1&1&0&0&0&1\\
1&0&0&1&0&1&1&1&0&1&1&1&0&0&1&0&1&0&0&0&0&1&1&1&0&0\\
1&0&0&0&1&0&1&1&0&0&1&1&0&0&0&0&1&1&0&0&0&0&0&1&0&0\\
0&1&1&0&1&1&1&1&1&0&0&1&1&0&0&0&0&0&0&0&1&0&1&0&1&1\\
 \end{tabular}
    
\end{table}

 \item[7)] Reordering every $26$ digits by Feistel inspired structure:
 \begin{table}[h!]
     \centering
     \begin{tabular}{p{0.2cm}p{0.2cm}p{0.2cm}p{0.2cm}p{0.2cm}p{0.2cm}p{0.2cm}p{0.2cm}p{0.2cm}p{0.2cm}p{0.2cm}p{0.2cm}p{0.2cm}p{0.2cm}p{0.2cm}p{0.2cm}p{0.2cm}p{0.2cm}p{0.2cm}p{0.2cm}p{0.2cm}p{0.2cm}p{0.2cm}p{0.2cm}p{0.2cm}p{0.2cm}}
0&0&0&1&1&0&1&0&0&1&1&0&1&1&1&1&0&0&0&0&0&0&0&1&1&1\\
0&0&1&1&0&1&0&0&0&1&0&0&0&1&0&1&0&1&1&0&0&0&0&0&1&1\\
1&1&0&1&0&0&1&0&1&0&1&1&0&1&0&0&0&0&0&0&1&1&0&1&0&0\\
1&1&0&0&0&1&0&1&1&0&0&0&1&0&0&0&1&0&1&0&0&1&1&0&0&1\\
0&1&0&1&0&0&0&0&1&1&1&0&0&1&1&0&0&0&1&1&1&1&0&0&1&0\\
0&0&0&1&1&0&0&0&0&0&1&0&0&1&0&0&1&0&0&1&1&0&0&0&1&0\\
0&0&0&0&0&0&0&1&0&1&0&1&1&0&1&1&0&1&1&1&0&1&1&0&0&0\\
\end{tabular}
 \end{table}
\end{itemize}
\vspace{0.2cm}
The sender sends this ciphertext to the receiver.\\
The receiver receives the ciphertext and runs receiver's side algorithm.
\subsection{Decryption}
\begin{itemize}
\item[1)] Get the ciphertext.

\begin{table}[h!]
    \centering
    \begin{tabular}{p{0.2cm}p{0.2cm}p{0.2cm}p{0.2cm}p{0.2cm}p{0.2cm}p{0.2cm}p{0.2cm}p{0.2cm}p{0.2cm}p{0.2cm}p{0.2cm}p{0.2cm}p{0.2cm}p{0.2cm}p{0.2cm}p{0.2cm}p{0.2cm}p{0.2cm}p{0.2cm}p{0.2cm}p{0.2cm}p{0.2cm}p{0.2cm}p{0.2cm}p{0.2cm}}
0&0&0&1&1&0&1&0&0&1&1&0&1&1&1&1&0&0&0&0&0&0&0&1&1&1\\
0&0&1&1&0&1&0&0&0&1&0&0&0&1&0&1&0&1&1&0&0&0&0&0&1&1\\
1&1&0&1&0&0&1&0&1&0&1&1&0&1&0&0&0&0&0&0&1&1&0&1&0&0\\
1&1&0&0&0&1&0&1&1&0&0&0&1&0&0&0&1&0&1&0&0&1&1&0&0&1\\
0&1&0&1&0&0&0&0&1&1&1&0&0&1&1&0&0&0&1&1&1&1&0&0&1&0\\
0&0&0&1&1&0&0&0&0&0&1&0&0&1&0&0&1&0&0&1&1&0&0&0&1&0\\
0&0&0&0&0&0&0&1&0&1&0&1&1&0&1&1&0&1&1&1&0&1&1&0&0&0\\
    \end{tabular}
   \end{table}
   \newpage
\item[2)] Apply a concept of diffusion by using two Feistel inspired scheme.
\begin{table}[h!]
    \centering
    \begin{tabular}{p{0.2cm}p{0.2cm}p{0.2cm}p{0.2cm}p{0.2cm}p{0.2cm}p{0.2cm}p{0.2cm}p{0.2cm}p{0.2cm}p{0.2cm}p{0.2cm}p{0.2cm}p{0.2cm}p{0.2cm}p{0.2cm}p{0.2cm}p{0.2cm}p{0.2cm}p{0.2cm}p{0.2cm}p{0.2cm}p{0.2cm}p{0.2cm}p{0.2cm}p{0.2cm}}
 1&1&1&1&1&0&1&0&0&1&0&1&0&0&0&0&1&1&0&1&0&0&1&1&0&1\\
 1&0&0&1&1&0&0&0&0&1&0&1&1&0&0&1&1&0&1&0&0&0&1&0&0&0\\
 0&1&0&1&0&0&1&1&0&0&0&1&0&1&1&0&1&0&0&1&0&1&0&1&1&0\\
 1&1&0&1&0&0&0&1&0&1&0&0&0&1&1&0&0&0&1&0&1&1&0&0&0&1\\
 1&0&0&1&0&1&1&1&0&1&1&1&0&0&1&0&1&0&0&0&0&1&1&1&0&0\\
 1&0&0&0&1&0&1&1&0&0&1&1&0&0&0&0&1&1&0&0&0&0&0&1&0&0\\
 0&1&1&0&1&1&1&1&1&0&0&1&1&0&0&0&0&0&0&0&1&0&1&0&1&1\\
    \end{tabular}
\end{table}

\item[3)] Perform XOR operation between the binary key and the Feistel decryption.
\begin{table}[h!]
    \centering
    \begin{tabular}{p{0.2cm}p{0.2cm}p{0.2cm}p{0.2cm}p{0.2cm}p{0.2cm}p{0.2cm}p{0.2cm}p{0.2cm}p{0.2cm}p{0.2cm}p{0.2cm}p{0.2cm}p{0.2cm}p{0.2cm}p{0.2cm}p{0.2cm}p{0.2cm}p{0.2cm}p{0.2cm}p{0.2cm}p{0.2cm}p{0.2cm}p{0.2cm}p{0.2cm}p{0.2cm}}
0&1&0&0&0&1&0&1&1&1&1&1&1&0&1&0&0&0&1&0&0&1&1&0&0&1\\
1&0&1&0&1&1&0&1&0&1&1&0&0&1&1&0&1&0&0&0&0&0&0&1&0&1\\
0&1&0&0&0&1&1&0&1&0&0&0&0&0&0&0&0&0&1&1&1&1&1&0&0&1\\
0&1&0&0&0&1&1&0&1&0&0&0&0&0&0&0&0&0&1&1&1&1&1&0&0&1\\
1&0&1&0&1&0&0&0&0&1&0&0&0&0&1&0&1&0&0&1&1&0&0&1&0&1\\
1&0&1&0&1&1&0&0&1&1&0&0&0&0&0&1&1&1&1&1&1&1&0&1&0&1\\
1&0&1&0&1&1&0&0&1&1&0&0&0&0&0&1&1&1&1&1&1&1&0&1&0&1\\

    \end{tabular}
\end{table}

\item[4)] Convert every $26$ binary cipher to corresponding decimal.
\begin{table}[h!]
    \centering
    \begin{tabular}{ccccccc}
18344089   &45455877& 18481401& 18481401& 44108389& 45287413& 45287413  \\
  
    \end{tabular}
    
\end{table}

\item[5)]  Decrypt each decimal by the Rabin decryption key.\\
\begin{table}[h!]
    \centering
    \begin{tabular}{cccc}
1018545 &  {\color{red}4283}& 46992172& 48006434  \\
47968600& 8210919& {\color{red}42117}& 39799798\\
{\color{red}4299}& 6346910& 48006418& 41663807\\
{\color{red}4299}& 6346910& 48006418& 41663807\\
13539284& 47968616& 34471433& {\color{red}42101}\\
47968602& 2875625& {\color{red}42115}& 45135092\\
47968602& 2875625& {\color{red}42115}& 45135092\\
\end{tabular}
\end{table}
 
\item[6)] Choose the number that starts with $42$ among each set of four outputs.
\begin{table}[h!]
    \centering
    \begin{tabular}{ccccccc}
    {\color{red}42}83& {\color{red}42}117& {\color{red}42}99& {\color{red}42}99& {\color{red}42}101& {\color{red}42}115& {\color{red}42}115\\
    \end{tabular}
\end{table}
  \item[7)] Remove the $42$ for each number.
\begin{table}[h!]
    \centering
    \begin{tabular}{ccccccc}
    83& 117& 99& 99& 101& 115& 115\\
    \end{tabular}
\end{table}

 \item[8)]  Convert each ASCII to its corresponding character.
\[ Success\]
  \end{itemize}

\section{Method Analysis}
Despite  many differences between DNA cryptography and conventional cryptography, both satisfy the same cryptographic requirements. The security of the proposed encryption  is totally dependent on three levels: The first level depends on the DNA OTP key. It is well known that the OTP is unconditionally secure if the key is truly random, has the specified length, is never completely or partially reused, and is kept in complete secrecy. Given these conditions, no adversary can obtain it. If a brute force attack is used to try to break the algorithm, the chances of finding the right combination are $1$ in $4^{(m\times26/2)}$, where $m$ is the message length. The likelihood of obtaining the correct combination for a message of length $10$ is therefore $1$ in $4^{130}=2^{260}$ combinations. It is a very remote possibility. The second level of security is the computational difficulty of factoring an integer of the Rabin key sharing system. The Rabin cryptosystem benefits from the fact that   the underlying problem has been shown to be equivalent to the integer factorization problem (\cite{15})  which currently is not known to be the case for the RSA problem, making the Rabin cryptosystem potentially more secure than the RSA (\cite{18}). The Rabin encryption method is more efficient in terms of computing squares $\mod n $ than the RSA, which needs to compute $e^{th}$ powers (\cite{1}). The Rabin cryptosystem is secure against specific plaintext attacks (\cite{18}) but it can be  cracked using chosen ciphertext attack, allowing the attacker to obtain the private key. The main flaw in the Rabin cryptosystem, which has kept it from being widely adopted in practice, is that decoding yields three false results in addition to the  true one, making it necessary to guess the true result (\cite{1}). Guessing is usually not hard if the plaintext represents a text message, however, if the plaintext is meant to represent a numerical value, the issue becomes one that needs to be solved via a disambiguation method. To solve this issue, redundancy in the message is required (three solutions to the problem are described in (\cite{15})), or alternatively, extra bits must be sent, one can choose plaintexts with unique structures or add padding. The third level of security is provided by the inclusion a Feistel-inspired structure in our suggested algorithm that improve the algorithm in terms of its security parameter (\cite{9}). We employ the Feistel-inspired structure for each of the $26$ digits. This makes things more complicated by adding some confusion and diffusion, which prevents the adversary from using any kind of brute force attack.

\section{Conclusion}
Applications of DNA cryptography are expanding quickly. The main approach of DNA cryptography is based on biochemical techniques for data encryption based on DNA sequences, which leads to  more effective new algorithms by integrating the characteristics of both biological and conventional cryptography (\cite{19}).  The goals of  DNA cryptography algorithms  are to increase data transmission security to the highest level possible and to decrease the computational complexity of encryption and decryption. In this work we propose a new encryption technique based on  DNA, Rabin cryptosystem, OTP, and Feistel inspired structure  to ensure higher security. Due of DNA's random nature (\cite{20}), the use of  DNA sequences  in OTP  makes the algorithm strong enough to fend off attacks. Given that DNA sequences are truly random and they are available on millions of websites, it is practically impossible to determine the random sequence utilized in encryption. 
The use the Rabin cryptosystem  potentially achieves a higher security  than RSA due to its   closer connection to the integer factorization problem. The idea of replacing the binary sequences  based on the Feistel structure also makes the proposed algorithm safer. Key sharing is a big disadvantage if the random binary key needs to be shared between the sender and the receiver but with the use of the method given in (\cite{16}), we have overcome this problem.  
Since the original data is never communicated in an open manner, the suggested method is more secure for data transfer over the internet.

\section{Acknowledgements}

\end{document}